\documentstyle[psfig]{cupconf}
\ifoldfss
\else
  \ifnfssone
    \newmathalphabet{\mathit}
      \addtoversion{normal}{\mathit}{cmr}{m}{it}
      \addtoversion{bold}{\mathit}{cmr}{bx}{it}
    \newmathalphabet{\mathcal}
      \addtoversion{normal}{\mathcal}{cmsy}{m}{n}
    \else
    \ifnfsstwo
    \fi
  \fi
\fi

\def\hexnumber#1{\ifcase#1 0\or1\or2\or3\or4\or5\or6\or7\or8\or9\or
 A\or B\or C\or D\or E\or F\fi }

\makeatletter
\ifx\CUP@mtlplain@loaded\undefined
\else
\fi
\makeatother
 \makeatletter
 \ifx\CUP@mtlplain@loaded\undefined
   \font\tenbmi=cmmib10 at 10pt
   \font\sevenbmi=cmmib10 at 7pt
   \font\fivebmi=cmmib10 at 5pt

   \newfam\bmifam
   \textfont\bmifam=\tenbmi
   \scriptfont\bmifam=\sevenbmi
   \scriptscriptfont\bmifam=\fivebmi
   
 \fi
 \makeatother
\ifnfsstwo

\fi
\ifnfssone

\fi
\ifoldfss

\fi

\mathchardef\varLambda="0103

\makeatletter
\ifx\CUP@mtlplain@loaded\undefined
\else
\fi
\makeatother
\makeatletter
\ifx\CUP@mtlplain@loaded\undefined
  \font\tenbms=cmbsy10
  \font\sevenbms=cmbsy10 at 7pt
  \font\fivebms=cmbsy10 at 5pt
  \newfam\bmsfam
  \textfont\bmsfam=\tenbms
  \scriptfont\bmsfam=\sevenbms
  \scriptscriptfont\bmsfam=\fivebms

  \edef\bsy@{\hexnumber\bmsfam}
  \mathchardef\bnabla="0\bsy@72
\fi
\makeatother
\def\etal{\mbox{\it et al.}}

\title[Conference Summary]{Where do we stand?}

\author[R.F.G.~Wyse]
{R\ls O\ls S\ls E\ls M\ls A\ls R\ls Y\ns F.\ns G.\ns W\ls Y\ls S\ls E}

\affiliation{Department of Physics \& Astronomy, The Johns Hopkins University, Baltimore, MD 21218
 USA}
\begin{document}
\ifnfssone
\else
  \ifnfsstwo
  \else
    \ifoldfss
      \let\mathcal\cal
      \let\mathrm\rm
      \let\mathsf\sf
    \fi
  \fi
\fi

\maketitle

\begin{abstract}

I review the understanding of bulges that emerged from the lively
discussions and presentations during the meeting, and emphasize areas
for future work.  The evidence is for a diversity of `bulges', and of 
formation mechanisms. 
\end{abstract}

\firstsection 
\section{What is a bulge?}

Classical bulges are  centrally-concentrated, high surface density,
three-dimensional stellar systems.  Their high density could arise
either because significant gaseous dissipation occurred during their
formation, or could simply reflect formation at very high
redshift (or some combination of these two, depending on the density). 
For illustration, equating the mean mass
density within the luminous parts of a galaxy (assumed to have circular velocity $v_c$ and radius $r_c$) with the cosmic mean
mass density at a given redshift, $z_f$, gives (e.g. Peebles 1989) 
$$z_f \sim 30 \, {1 \over f_c \Omega^{1/3}}\, ({v_c \over 250 {\rm
km/s}})^{2/3}\, ({10 {\rm kpc} \over r_c})^{2/3},$$ where $f_c$ is the
collapse factor of the proto-galaxy, being at least the factor 2 of
dissipationless collapse, and probably higher so that bulges, as
observed, are self-gravitating, meaning that they have collapsed
relative to their dark haloes.

The majority view at the meeting, consistent with the observations, is
that indeed proto-bulges radiated away binding energy, but also at
least their stars formed at relatively high redshift.  One must always
be careful to distinguish between the epoch at which the stars now in
a bulge formed, and the epoch of formation of the bulge system itself
(as emphasized by Pfenniger, this volume).  Of course if the bulge
formed with significant dissipation, meaning gas physics dominated,
then the star formation and bulge formation probably occured together.

The small length-scale of bulges, combined with their modest rotation
velocity, leads to a low value of their angular momentum per unit
mass.  Indeed, in the Milky Way Galaxy, the angular momentum
distribution of the bulge is similar to that of the slowly-rotating
stellar halo, and different from that of the disk, strongly suggestive
of a bulge--halo connection, perhaps via gas ejection from halo
star-forming regions (e.g. Wyse \& Gilmore 1992). One can appeal to
bulges forming from the low angular momentum regions of the
proto-galaxy, a variant on the Eggen, Lynden-Bell \& Sandage (1967)
`monolithic collapse' scenario,  explored further by van den Bosch
(1998 and this volume). Or one can posit angular momentum transport
prior to the formation of the bulge, taking angular momentum away from
the central regions, and depositing it in the outer regions.
Such transport of angular momentum could perhaps occur
during hierarchical merging, by dynamical friction and gravitational
torques, although
one must be careful not to end up with too small a disk due to over-efficient angular momentum re-arrangement 
(e.g. Zurek, Quinn \& Salmon 1988; Navarro \& Benz 1991; Navarro \& Steinmetz 1997).  More modest amounts of angular momentum transport may be achieved by  
some viscosity in the early disk (e.g. Zhang \& Wyse 1999).

A recurring theme of the meeting was that large bulges (of early-type
disk galaxies?) are related to ellipticals while small bulges
(intermediate--late-type disk galaxies?) are more closely allied to
disks.  We need to be very clear about the observational 
selection criteria used in the 
definition of samples, and how this could bias our conclusions.  As we will
see below, the Milky Way bulge shows characteristics of {\it both\/}
early- and late-type bulges, and will feature in {\it both\/}
bulge--elliptical connections and bulge--disk connections.

\subsection{The elliptical--bulge connection}

There has been remarkably little new kinematic data for representative
samples of bulges (as opposed to detailed study of particular
individual bulges, chosen for their unusual characteristics) since the
pioneering work of the 1970s and 1980s.  As demonstrated by Davies
\etal\/ (1983), the bulges of early-type spirals are like ellipticals
of equal luminosity in terms of rotational support, and are consistent
with being isotropic oblate rotators i.e. with having an isotropic
stellar velocity dispersion tensor, and being flattened by rotation
about their minor axis.  This sample was biased towards early-type
spirals to facilitate bulge--disk decomposition, by observing edge-on
systems with a prominent bulge.  The bulge of the Milky Way Galaxy can
be observed to match the techniques employed in the study of the
bulges of external galaxies and, also then has stellar kinematics
consistent with being an isotropic rotator (Ibata \& Gilmore 1995a,b;
Minniti 1996), as shown in Figure 1 here.  

\begin{figure} 
\hskip0.5in
\psfig{file=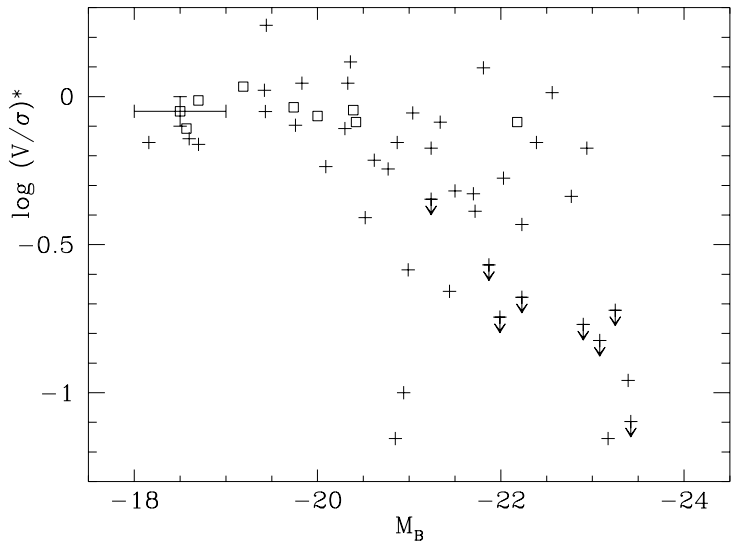,width=4in,angle=270} 
\caption{The level of rotational support
  as measured by (V/$\sigma$)*, which has the value unity for an
  isotropic oblate rotator, against absolute magnitude for elliptical
  galaxies (plus symbols) and bulges of early-type 
spirals (open squares); data
  from Davies {\it et al.} 1983.  The bulge of the Milky Way, with
kinematic  quantities and flattening estimated in a similar manner as 
for the external galaxies, is indicated by
  the point with error bars.}
\end{figure} 

The trend apparent in Figure~1, and discussed more fully in Davies
\etal\/ (1983), is that the level of rotational support in ellipticals
increases as the luminosity of the elliptical decreases.  The surface
brightness of ellipticals also increases with decreasing luminosity,
at least down to the luminosity of M32 (the dwarf spheroidal galaxies
are another matter), as noted by Kormendy (1977), Wirth \& Gallagher
(1984) and many subsequent papers.  These two relations are consistent
with an increasing level of importance of dissipation in ellipticals 
with decreasing galaxy 
luminosity (Wyse \& Jones 1984).  

Further, the bulges of S0-Sc disk galaxies follow the general trend of  
the Kormendy (1977) 
relations, in that smaller bulges are denser (de Jong 1996; 
Carlberg, this volume; see Figure~3 below for details).  
Thus one interpretation of Figure 1 is then that (some) bulges 
too formed with significant dissipation. 

As discussed by several speakers, the bulges of S0-Sc disk galaxies
have approximately the same Mg2 -- velocity dispersion relation as do
ellipticals (Jablonka \etal\/ 1996; Idiart \etal\/ 1997; see Renzini
this volume), although the actual physics behind this correlation is
not uniquely constrained.  The properties of line-strength gradients
in ellipticals of a range of luminosities are consistent with lower
luminosity ellipticals forming with more dissipation than the more
luminous ellipticals (Carollo, Danziger \& Buson 1993).  Again these
results are suggestive that bulges are similar to low-luminosity
ellipticals, and that gas dissipation was important.

The detailed interpretation of the line-strength data in terms of the
actual age and metallicity distributions of the stars is extremely
complex and as yet no definitive statements can be made.  There is a
clear need for more data, including radial gradients, and for more
models (see Trager, this volume).  The broad-band colors of (some)
bulges are consistent with those of the stellar populations in
early-type galaxies in the Coma cluster (Peletier \& Davies, this
volume). We still need better models to interpret even broad-band
colors.

\subsection{The disk--bulge connection}

The surface-brightness profiles of bulges in later-type disk galaxies
are better fit by an exponential law than by the steeper de Vaucouleurs
profile, which in turn is a better fit for the bulges of early-type disk
galaxies (Andredakis, Peletier \& Balcells 1995; de Jong 1996).  
The sizes of bulges are statistically related
to those of the disks in which they are embedded, and indeed the
(exponential) scale-lengths of bulges are around one-tenth that of
their disk; this correlation is better for late-type spirals than for
early types (Courteau, de Jong \& Broeils 1996).  
The projected starlight of the bulge
of the Milky Way can be reasonably well-approximated by exponentials
(vertically and in the plane); the Milky Way then fits within the
scatter of the correlation of the external galaxies.

The optical colors of bulges are approximately the same as those of
the inner disk, for the range of Hubble types S0-Sd (Balcells \&
Peletier 1994; de Jong 1996), but as ever the decomposition of the
light curves is difficult, as is correction for dust.  This
correlation implies similar stellar populations in bulges and their
disks, as may be expected if bulges form from their disks (see
Pfenniger, this volume).  Thus, should there be a variation of mean
stellar age from disk to disk, as may be expected from the range of
colors observed, and indeed from observations of gas fraction etc.,
together with models of star formation in disks, one would expect a
corresponding range in the mean stellar age of the different bulges.
However, Peletier \& Davies (this volume) find only a narrow range in
bulge ages for their sample, based on optical--IR colors.  More data
are clearly needed.

Figure~1 demonstrated the similarity in their kinematics between
bulges and ellipticals of the same luminosity; Figure~2 (taken from
Franx 1993) illustrates some of the complexity of bulge kinematics,
and emphasizes the need to be aware of the selection criteria -- not
all bulges are the same.  The left-hand panel shows that in terms of
the ratio of stellar velocity dispersion to true circular velocity
(not the rotational streaming velocity), bulges scatter below
ellipticals.  Further, the right-hand panel shows that bulges of
late-type disk galaxies have values of this ratio similar to that
typical of inner disks (from Bottema 1993).  The Milky Way bulge in
this plot is quite typical ($\sigma/{\rm V_c} \sim 0.5$, B/T$ \sim
0.25$).

\begin{figure}
\vskip -2.75truein
\psfig{file=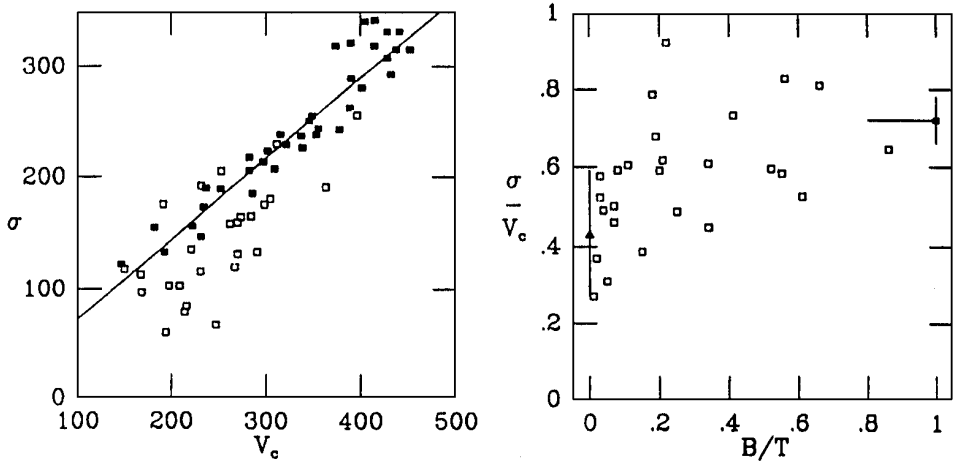,width=5.75in}
\vskip -2.75 truein
\caption{{\it a}) The central velocity dispersion of stellar tracers,
$\sigma$, against dark halo circular velocity, $v_c$. Open symbols represent 
bulges; closed symbols represent ellipticals. Circular velocities for the
ellipticals are derived from models, as described by Franx (1993). ({\it b})
The ratio of velocity dispersion in the bulge to dark halo circular
velocity, $\sigma/v_c$, taken from Franx (1993), plotted as a function
of bulge-to-total luminosity (B/T) ratio, for the entire 
range of Hubble Type. The triangle at left is valid for the inner 
regions of pure disks, the
square at right for ellipticals. Note that systems with low B/T have
kinematics almost equal to those of inner disks.}
\end{figure} 

Complexity in the relationship between surface brightness and
scale-length for bulges is illustrated in Figure~3, based on {\tt
WFPC2} data from Carollo (1999).  The plot shows that while the large,
$R^{1/4}$-law bulges follow the same scaling as ellipticals, the
smaller, exponential-profile bulges are offset to lower surface
brightnesses and occupy the extension to smaller scalelengths (by
about a factor of ten, as noted above) of the locus of late-type
disks.  This strengthens the disk--bulge connection for these small
bulges.  However, Carollo (1999) finds both R$^{1/4}$ and exponential bulges 
in apparently very similar disks, so some additional parameter is important.

\begin{figure}
\vskip -1.5truein
\hskip  -0.3truein
\psfig{file=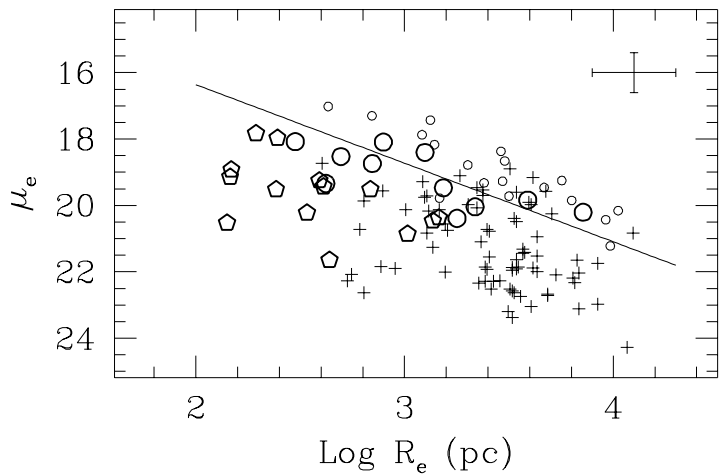,width=5.75in}
\vskip -1.25 truein
\caption{The mean $V$-band  surface brightness $\mu_e$ within the half-light
radius $R_e$, as a function of log$R_e$ (in pc). The {\tt WFPC2}
measurements are shown with  pentagons for the exponential
bulges and large circles for the `classical' $R^{1/4}$-law bulges.
Comparison data from the literature are  shown for the $R^{1/4}$
bulges from Bender \etal\/ (1992; small circles) and  Scd-Sm disks from Burstein \etal\/ (1997; crosses).  
The solid 
line is the best fit to the elliptical galaxy sequence (data from
Bender \etal\/ 1992 and from Burstein \etal\/ 1987).  
The typical 1-$\sigma$ error bar for the {\tt WFPC2}
masurements is shown in the upper-right corner.}
\end{figure}

Association of `peanut' bulges with bars, which are essentially a disk
phenomenon, was made in several contributions, using both gas and
stellar kinematics (Kuijken; Bureau). However, the pronounced `peanut'
in the early COBE images of the Milky Way was apparently largely an
artefact of patchy dust, and the amplitude of such a morphology in the
bulge of the Milky Way is not reliably established (Binney, Gerhard \& Spergel 
1997).  As emphasized by Pfenniger (this volume), the kinematical and
dynamical effects of bars are 3-dimensional; they can scatter stars by
resonances, and/or themselves go unstable, fatten and dissolve,
leading to a bulge. Which process dominates?  There is a wealth of
fascinating physics to explore.  The modellers need to make more
contact with observations, including predictions for direct comparison
with the stellar kinematics, ages of stars, surface brightness
profiles etc.

M33 has neither a bulge nor a bar, but does have a central nucleus,
and of course a substantial disk.  Such systems need to be discussed
in this context.  The central nucleus of the Milky Way contains a
black hole and star clusters of mass fraction well below the 1\% or so
estimated to destroy a bar (Norman, Sellwood \& Hasan  1996), if we
associate all the $10^{10}$M$_\odot$ of the bulge with the bar. Indeed
it is somewhat of a curiosity that the Milky Way does not fit the
relationship between black hole mass and bulge mass found by
Magorrian \etal\/ (1998).

\section{When do Bulges form?}

The fossil evidence from Local Group galaxies constrains  the ages of
the stars presently in bulges, which could be rather different from
the age of the morphological system.

The overwhelming evidence (contributions by Gilmore, Frogel, Renzini,
Rich) for the Milky Way bulge is that its stars are old, except for a
very small scaleheight young component -- and since all components of
the Galaxy have their peak surface brightnesses in the
center,\footnote{Some of the decompositions of the COBE data have
modelled the disk with a hole in the central regions, which I 
believe points to continuing uncertainly in the interpretation
of those data in terms of the parameterizations of the different
components along the line-of-sight.}  this is as likely to be
associated with the disk.  Further, as discussed by Rich (this volume)
the dominant stellar population even in the nuclear regions is
apparently old.

The situation in external bulges in the local universe is more
uncertain, but is consistent with stars in large bulges being `old', which
means forming perhaps 10Gyr ago.

Direct studies of morphology at high redshift require HST and are at
present based on small samples and must be treated with caution if
attempting to draw general conclusions.  The Hubble Deep Field (HDF)
has provided much of the field galaxy sample (as opposed to members of
galaxy clusters).  Recently, Abraham \etal\/ (1999a) analysed the
spatially-resolved colors of galaxies of known redshift in the HDF.
In contrast to the case of cluster ellipticals discussed by Renzini
(this volume), they find that almost half of their (small) sample of
field ellipticals at intermediate redshift ($0.4 < z < 1$) show
evidence for a range of stellar ages. The color gradients in the
galaxies for which they could derive a reasonable bulge--disk
decomposition are consistent with the mean stellar ages of the bulges
being older than those of their disks.  These
authors argue that this presents difficulties for secular evolution
models, but again one must remember the possible selection biases.  
Abraham \etal\/ (1999b) further find a significant deficit of barred
galaxies for redshifts above 0.5; as those authors note, more data for
a wider range of rest-frame colors and redshifts 
are needed confirm this result, and then 
to decide on a robust
interpretation.  As discussed by Lilly (these proceedings), there is
strong evidence from SCUBA data for the existence of compact galaxies
with high star-formation rates at high redshift, consistent with
proto-spheroids forming in a starburst.

The age distribution of inner disks is of obvious importance for
constraining scenarios of disk--bulge formation.  Unfortunately, we do
not know this well, even in the Milky Way.  Indeed, we do not have a
good understanding of the star formation history even at the solar
neighborhood. We do know that out to a few kpc from the Sun there are
stars in the thin and thick disks that are as old as the globular
clusters (Edvardsson \etal\/ 1993; Gilmore, Wyse \& Jones 1995).  The
stellar color--magnitude data, the chemical abundances and the white
dwarf luminosity function data are all broadly consistent with a local
(solar neighborhood) star formation rate that has been approximately
constant, back to $\sim 12$Gyr (e.g. Noh \& Scalo 1990; Rocha-Pinto \&
Maciel 1997).  Most models of star formation in disks predict that the
central regions should evolve faster, and hence the mean stellar age
should be older in the inner disk than in the outer disk. Thus perhaps
indeed predominantly-old bulges can be formed  recently, from old
stars in the central parts of disks.  
But one really has to be careful to avoid a
significant age range in the bulge, reflecting 
the continuing star formation in the
disk up to the time of bulge formation.

Simulations of hierarchical clustering galaxy formation predict
`bulges' to form stars at redshift of $z \sim 2$ (peak) even if
assembled later (Frenk, oral presentation this meeting; Baugh, Cole,
Frenk \& Lacey 1998).  In these scenarios, bulges (and ellipticals)
form from mergers between pre-existing disk galaxies, and consist of a
mix of the disk stars, plus, in some versions, new star formation in
the central regions resulting from the disk gas being driven there
during the merger. Disks are then (re-)accreted around these
bulges. Thus bulges in galaxies with relatively big disks (i.e. Scs)
should be the oldest bulges, and bulges with small disks should be the
youngest (Baugh, Cole \& Frenk 1996; Kauffmann 1996).  This is not
obviously consistent with the observations presented at the meeting.

A preliminary attempt to make detailed predictions and see if the `bulges' 
in these models fit the observed scaling between size and luminosity
was presented at the meeting by Lacey; the models did not include
dissipation and failed to produce small enough bulges.  This is
further evidence that the high phase space densities of bulges require
dissipation (cf. Wyse 1998). 

\section{What are the Timescales -- duration of bulge formation?}

The finest time-resolution in studies of stellar populations is available 
from study of the patterns of
elemental abundances in individual stars, as discussed in this 
volume by Renzini; the elemental signature of a short duration
of star formation is a pattern of enhanced
$\alpha$-elements as produced by Type~II supernovae alone only.  
The bulge of the Milky Way is surprisingly under-studied and 
really do need more
data for field stars; for the extant small sample,
different $\alpha$-elements show different patterns (McWilliam \& Rich
1994), unexplained within the context of solely Type~II supernovae yields
(e.g. Worthey 1998) or in comparison with the element ratios of 
stars in the stellar halo.  It is worth
noting however that the elemental abundances seen in the bulge field stars and
in the bulge (or thick-disk?) globular clusters are consistent with a
normal massive-star stellar IMF (cf. Wyse \& Gilmore 1992), 
as also seen via star counts for the lower mass stars in Baade's
window (Holtzman \etal\/ 1998).

Color-magnitude diagrams of old populations can only constrain the
duration of star formation to be less than many Gyr, due to the
crowding of the isochrones (reflecting the long main-sequence
lifetimes of low-mass stars). Further, one needs to know the
metallicity distributions, and crucially for the Milky Way bulge, the
distance distribution, since foreground disk stars are a difficult
contaminant.

As mentioned above, hierarchical-clustering and merging scenarios
predict a many Gyr spread in ages of bulge stars, but we need a better
quantification of `many'.  And again, a significant age spread is
predicted in the simpler secular evolution models, although the
restriction to only one early disk--bar--bulge episode would minimise
it.

The shortest durations of star formation are predicted by the 
starburst models wherein pre-assembled gas forms stars on only a few 
free-fall times, but the physics of the assembly of the gas will also play a role (Carlberg, this volume, who favors wind-regulated 
accretion of gas-rich satellites; Elmegreen, this volume, who favors 
unregulated, monolithic collapse). That very high star formation rates 
happened in some systems at high redshift is supported by the SCUBA 
observations (Lilly, this volume), but important aspects of the model 
obviously need to be worked out (e.g. is there or isn't there a 
dominant supernova-driven wind?)

\section{Constraints from Physical Properties}

\subsection{Angular momentum distributions} 

The hierarchical-clustering and merging scenario predicts misalignment
in the angular momentum vector of different shells of material around
a peak.  This may be expected to translate into some persistent
misalignment between disk and bulge, and even counter-rotating
components.  While examples of such systems exist (see Bertola \etal\/
this volume), these would appear to be the exception rather than the rule
(see Kuijken, this volume).  

Quantification of the specific angular momemtum distributions of disks
and bulges is obviously desirable, but the observational
determinations are dependent on not only detailed kinematic data, but
also the decomposition of the light profile (and M/L).  Note that 
in the Milky Way  the determination of the kinematic properties of the
bulge -- and in particular any gradients -- requires very careful treatment of contamination by the disk (see Ibata \& Gilmore
1995a,b; Tiede \& Terndrup 1997 for details). The extant 
theoretical predictions of angular momentum distributions of bulge and of disk 
are also not sufficient.

\subsection{Central star clusters and bars} 

`Secular evolution' models for forming bulges from inner disks naively
predict an anti-correlation between significant central mass
concentrations and bars, since in these models the clusters destroy
the bar.  There is a particular need to determine how many cycles of
bar formation/dissolution are expected theoretically, and how many are
allowed by the observations.  The uniform old age of bulges, including
that of the Milky Way, suggested by most of the evidence presented at this meeting (but again remember possible selection effects) argues
strongly for only one such episode, and as noted by Gilmore (this
volume), the disk must still continue into the central regions.  The
relative frequency of bars, exponential versus R$^{1/4}$ bulges,
central star clusters etc.\/ is as yet poorly quantified.  The initial
results of an HST {\tt WFPC2} and {\tt NICMOS} imaging survey of
nearby spiral galaxies (Carollo 1999) have revealed some of the
complexity of the inner regions of these systems, finding a high
fraction of photometrically-distinct compact sources sitting at the
galactic centers. These `nuclei' have surface brightnesses and radii
ranging from those typical of the old Milky Way globular clusters to
those of the young star-clusters found in interacting galaxies
(e.g. Whitmore \etal\ 1993; Whitmore \& Schweizer 1995), with typical
half-light radii of a few pc up to $\approx 20$pc.  Many of the nuclei
are embedded in bulge-less disks or in bulge-like structures whose
light distribution is too dusty/star-forming to be meaningfully
modelled.  Every exponential bulge was found to contain a nucleus, and
further the luminosity of the nucleus was consistent with its being 
sufficiently massive to have destroyed a bar of the same mass as the (exponential) bulge.  Are these nuclei the central mass concentrations of the models?

The ${\rm V-H}$ color distribution of the exponential bulges is rather
broad, and peaks at ${\rm V-H \sim 0.96}$, significantly bluer, by
about 0.4~mag, than the value typical of the R$^{1/4}$ bulges (Carollo \etal\/ 1999). If this
bluer color can be ascribed to a younger age, this would indicate that
exponential bulges are the preferred mode, for bulges forming more
recently.  The relatively massive central clusters
found in these  exponential bulges could theoretically
prevent subsequent  bar formation, and removing  
the possibility of successive cycles of bar
formation -- gas inflow -- formation of central object -- bar 
dissolution mechanism (as was discussed also by Rix in his oral presentation at
this meeting).

\subsection{Chemical abundance distributions}

The K-giants in the Milky Way bulge have a very broad metallicity
distribution, both in Baade's window (at a projected Galactocentric
distance of around 500pc; Rich 1988) and at projected distances of
several kpc (Ibata \& Gilmore 1995a,b).  The breadth of the
metallicity distribution in the bulge contrasts with that narrow
distribution observed in the disk at the solar neighborhood. The lack
of metal-poor stars in the local disk conflicts with predictions of
the `simple model' of chemical evolution, and is the famous `G-dwarf
problem'.  
One hastens to add that the fact the Milky Way bulge has a broad
distribution, and indeed fits the predictions of the `simple model' of
chemical evolution, does not mean that any or all of the many
assumptions of the `simple model' are valid; another example of a
stellar system with a metallicity distribution that is well-fit by the
simple model (albeit with a reduced yield) is the stellar halo of the
Milky Way. The G-dwarf problems has many solutions, the most popular
of which is to postulate gas inflows (e.g. Tinsley 1980).  The width
of a metallicity distribution is related to the ratio of inflow time
to star formation time, and perhaps the wider metallicity distribution
in the bulge can be interpreted in terms of very rapid star formation,
occuring too fast for inflow to affect the metallicity structure.

The M-giants studied by Frogel (this volume) in the inner 100pc or so
of the bulge do appear to have a narrow metallicity distribution, but
this may reflect the bias inherent in the sample selection by such a
late spectral type; data for K-giants are desirable, both because they
are a more representative evolutionary phase of low-mass stars, and
because their spectra are easier to interpret and use to determine
metallicities, than are M-giants.

From the width of the giant branch in color-magnitude diagrams, the
M31 bulge is inferred to have a rather broad metallicity distribution
in its outer parts, but a narrow metallicity distribution interior to
1kpc (Renzini, this volume; Rich, this volume).  Perhaps this
variation in width also reflects a variation of the ratio of
star-formation rate to gas inflow rate, this time a variation with
radius within the bulge. At face value, the opposite trend -- one with
a broader metallicity distribution in the inner, more dense parts --
may be expected in models where the local star formation rate is
determined by a non-linear function of gas density, but the flow rate
is given by the inverse of the dynamical time (proportional to the
square root of density), so that the ratio of star formation time to
flow time decreases with increasing density.

The stellar populations of the resolved bulges in the Local Group are
not compatible with their formation via accretion and assimilation of
satellites and or globulars like those remaining today -- the bulges
are too metal-rich, and have too narrow an age distribution.  However,
perhaps some part of the metal-poor tail in the Milky Way bulge could
be due to accretion of the dense, metal-poor and old globular
clusters.  Note that for stellar satellite systems with a realistic density
profile, a significant fraction of the stars will be tidally removed
far out in the halo, and only a fraction will make it into the center
(Syer and White 1998; see Kuijken, this volume).  Kuijken (this
volume) notes that the timescale of satellite accretion is rather
long, so that any bulge-building by this means should be on-going.
This raises a further difficulty, in that the old, metal-rich bulge
stars are unlike those in typical satellites. A graphic illustration
of the difference in stellar populations between the bulge of the
Milky Way and the Sagittarius dwarf, one of the more massive satellite
galaxies of the Milky Way, is shown in Figure~4.

\begin{figure} 
\vskip -3.75in
\psfig{file=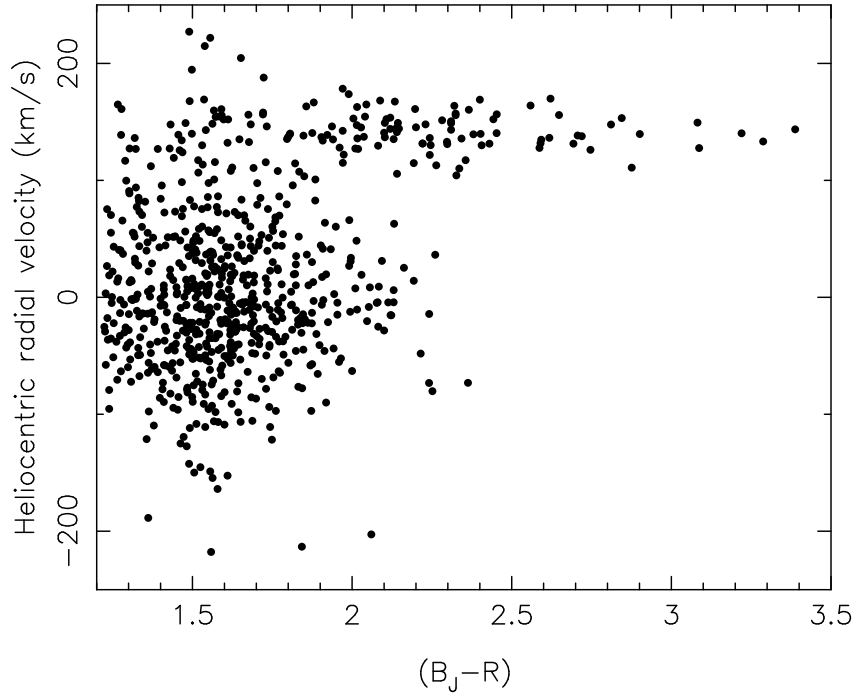,width=6in}

  \caption{Heliocentric radial velocities of the sample of K-giant
stars observed by Ibata, Gilmore \& Irwin (1994) in lines-of-sight
towards the Milky Way bulge ($\ell = -5^{\rm o}, \, b = -12^{\rm o},
\, -15^{\rm o}, \, -20^{\rm o}$).  Note the narrow velocity-dispersion
subsample centered at around 150km/s.  These stars are members of the
Sagittarius dwarf spheroidal galaxy, which was discovered from these
data.  The reddest stars are exclusively members of this galaxy,
illustrating the real difference in stellar populations between the
bulge and this satellite galaxy. The bulge cannot have formed from a
simple merger of satellites like the Sagittarius dwarf. }
\end{figure} 

\subsection{Chemical Abundance Gradients}

A strong chemical abundance gradient is a signature of slow,
dissipative collapse.  Such gradients are weakened, but not erased, by
any subsequent mergers (e.g. White 1980; Barnes \& Hernquist 1992).
There are no clear predictions for secular evolution models (but are
needed).

Observationally, there are weak or minimal amplitude gradients in mean
metallicity in resolved bulges (Milky Way Galaxy -- Gilmore, Frogel,
this volume; M31 -- Frogel, Rich, Renzini, this volume).  As
mentioned, the interpretation of absorption line-strengths remains
ambiguous, and we need more data and models.

\section{Summary}

Bulges are diverse in their properties, and probably in their
formation mechanisms, or at least in the dominant physics at the
epochs of star formation and/or assembly.  Perhaps the differences are
just a matter of degree, since, for example, even `monolithic
collapse' involves fragmentation, with subsequent star formation in
the fragments.  A centrally-concentrated profile appears to match
`maximum entropy' arguments (Tremaine, Henon \&  Lynden-Bell 1986) for
the end-point of violent relaxation of a cold, clumpy system,
independently of the details of the evolution to that end-point.

The overall trends of the observations are that small bulges, of
late-type disk galaxies, show a strong connection to their disk, while
big bulges, of early-type disk galaxies, are more like the low-luminsity
extension of the elliptical galaxy sequence.
The bulge of the Milky Way appears to straddle these two generalities,
having an affinity for its disk in terms of structure, but having the
old, metal-rich population associated with `spheroids'.

What does this mean?  Even the casual reader should have noted the
not-infrequent occurrence of the sentiment `more data and models are
needed' in the text above.  We are at the stage of requiring robust
quantitative results from both theory and observations. 

More specifically, for the Milky Way, we require good HST
color-magnitude diagrams for more lines-of-sight towards the Milky Way
bulge, following the work of Feltzing \& Gilmore (1999) in
establishing the association of a younger stellar population with
foreground disk. We also require good reddening maps and metallicity
data to aid the interpretation of these color-magnitude diagrams.  The
inner disk of the Milky Way is remarkably under-studied, and again age
and metallicity distributions -- and stellar kinematics -- are
obviously crucial in determining the similarity or otherwise of inner
disk and bulge.  Further, we need to understand the relationship
between the `bulge' globular clusters and the bulge field population;
present models of globular-cluster formation appeal to pre-enrichment
to provide the uniform enrichment within a given cluster, so it is not
obvious that the enrichment signatures of cluster stars and field
stars should be the same.  Elemental abundances for 
statistically-significant samples of unbiased tracers of the 
field in a variety of
lines-of-sight are required to understand the history of star
formation.  

A combination of HST and ground-based (to probe both small- and
large-scale structure) broad-band optical and IR colors, and surface
brightness profiles, are still lacking for large samples, including
the whole range of spiral Hubble types.  These data should allow a
robust quantification of the correlations between morphologies.
Basic kinematic data, including gradients, should be obtained for a
representative sample of bulges and disks.  While we may lack the
means at present for a unique interpretation of absorption
line-strength data, the straightforward test for continuity 
in the line strengths  from bulges to their disks is meaningful.

The redshift of statistically-significant samples of galaxies 
is being continually pushed back (at what point will this pose a 
real problem for CDM?) and HST and the next generation of telescopes 
should provide robust morphological classifications.  
We will no doubt see evolution, but need to have the model 
predictions to be able to distinguish the underlying physics behind 
the evolution.  

`Secular-evolution' models are their early stages of development, but
several key questions may be posed. While it may be reasonable to
comment that a correlation between bulge scale-length and disk
scale-length points to a connection between bulge and disk, can the
models `post'-dict the factor of ten that is observed?  Can they
predict the frequency with which one should see barred spirals today,
even ones with big bulges? Are all bars the same?  Are there too many
bars and/or central concentrations 
observed for the models of bar dissolution? Or is the dominant
mechanism of bulge-building in this scenario actually scattering of
disk stars through resonant coupling, rather than bar dissolution?
How can this be compatible with uniformly old bulges?  But are
exponential bulges (apart from the Milky Way bulge) composed of old
stars? 

Cold-dark-matter dominated cosmologies gained popularity partially
because of their robust predictive power, a requirement for a good
theory, in terms of the large-scale structure formed by the
dissipationless dark haloes, (e.g. Davis, Efstathiou, Frenk \& White
1992).  The predictions for the luminous components, the galaxies as
we observe them, have not yet achieved the same level of maturity. Advocates of
merging and hierarchical clustering should quantify further the ages
of stars now in bulges, and the epoch of assembly into bulges.  What
is predicted for the age spread within a typical bulge like the Milky
Way?  What fraction of bulges should have angular momentum vector
misaligned with their disk?  Should colors of bulge and disk be
correlated?

If bulges form in a `star-burst', what is the role of a
supernova-driven wind? In this context, the X-ray properties of
bulges, including the Milky Way, should constrain the ability of the
bulge potential well to retain hot gas.

Where do we stand? -- inspired to get to work!

\begin{acknowledgments}

I acknowledge support from  NASA ATP grant NAG5-3928.

\end{acknowledgments}

\end{document}